\documentclass[aps,prl,twocolumn,showpacs,superscriptaddress,citeautoscript,floatfix,longbibliography]{revtex4-1}
\usepackage{amsfonts}
\usepackage{amsmath}
\usepackage{amssymb}
\usepackage{graphicx}
\usepackage{dcolumn}
\usepackage{bm}
\usepackage[svgnames]{xcolor}
\usepackage[
	colorlinks=True,linkcolor=DarkRed,citecolor=ForestGreen,urlcolor=MediumBlue,
	pdfstartview=FitH,bookmarks=False,pdfpagemode=UseNone
]{hyperref}

\begin{document}
\title{Attractive interaction between superconducting vortices in tilted magnetic fields}
\author{Alexandre Correa}
\affiliation{Instituto de Ciencia de Materiales de Madrid, Consejo Superior de
Investigaciones Cient\'{\i}ficas (ICMM-CSIC), Sor Juana In\'es de la Cruz 3,
28049 Madrid, Spain}
\author{Federico Mompe\'an}
\affiliation{Instituto de Ciencia de Materiales de Madrid, Consejo Superior de
Investigaciones Cient\'{\i}ficas (ICMM-CSIC), Sor Juana In\'es de la Cruz 3,
28049 Madrid, Spain}
\affiliation{Unidad Asociada de Bajas Temperaturas y Altos Campos Magn\'eticos, UAM, CSIC,
Cantoblanco, E-28049 Madrid, Spain}
\author{Isabel Guillam\'on}
\affiliation{Unidad Asociada de Bajas Temperaturas y Altos Campos Magn\'eticos, UAM, CSIC,
Cantoblanco, E-28049 Madrid, Spain}
\affiliation{Laboratorio de Bajas Temperaturas, Departamento de F\'isica de la Materia
Condensada, Instituto Nicol\'as Cabrera and Condensed Matter Physics Center (IFIMAC),
Universidad Aut\'onoma de Madrid, E-28049 Madrid, Spain}
\author{Edwin Herrera}
\affiliation{Unidad Asociada de Bajas Temperaturas y Altos Campos Magn\'eticos, UAM, CSIC,
Cantoblanco, E-28049 Madrid, Spain}
\affiliation{Laboratorio de Bajas Temperaturas, Departamento de F\'isica de la Materia
Condensada, Instituto Nicol\'as Cabrera and Condensed Matter Physics Center (IFIMAC),
Universidad Aut\'onoma de Madrid, E-28049 Madrid, Spain}
\affiliation{Facultad de ingenieria y ciencias b\'asicas, Universidad Central, Bogot\'a 110311, Colombia}
\author{Mar Garc{\'i}a-Hern{\'a}ndez}
\affiliation{Instituto de Ciencia de Materiales de Madrid, Consejo Superior de
Investigaciones Cient\'{\i}ficas (ICMM-CSIC), Sor Juana In\'es de la Cruz 3,
28049 Madrid, Spain}
\affiliation{Unidad Asociada de Bajas Temperaturas y Altos Campos Magn\'eticos, UAM, CSIC,
Cantoblanco, E-28049 Madrid, Spain}
\author{Takashi Yamamoto}
\affiliation{QuTech, Delft University of Technology, PO Box 5046, 2600 GA, Delft, The Netherlands}
\author{Takanari Kashiwagi}
\affiliation{Graduate School of Pure and Applied Sciences, University of Tsukuba, Tennodai,
Tsukuba Ibaraki, 305-8573, Japan}
\author{Kazuo Kadowaki}
\affiliation{Graduate School of Pure and Applied Sciences, University of Tsukuba, Tennodai,
Tsukuba Ibaraki, 305-8573, Japan}
\author{Alexander I. Buzdin}
\affiliation{Condensed Matter Theory Group, LOMA, UMR 5798, University of Bordeaux, F-33405 Talence, France}
\author{Hermann Suderow}
\affiliation{Unidad Asociada de Bajas Temperaturas y Altos Campos Magn\'eticos, UAM, CSIC,
Cantoblanco, E-28049 Madrid, Spain}
\affiliation{Laboratorio de Bajas Temperaturas, Departamento de F\'isica de la Materia
Condensada, Instituto Nicol\'as Cabrera and Condensed Matter Physics Center (IFIMAC),
Universidad Aut\'onoma de Madrid, E-28049 Madrid, Spain}
\author{Carmen Munuera}
\affiliation{Instituto de Ciencia de Materiales de Madrid, Consejo Superior de
Investigaciones Cient\'{\i}ficas (ICMM-CSIC), Sor Juana In\'es de la Cruz 3,
28049 Madrid, Spain}
\affiliation{Unidad Asociada de Bajas Temperaturas y Altos Campos Magn\'eticos, UAM, CSIC,
Cantoblanco, E-28049 Madrid, Spain}

\begin{abstract}
Many practical applications of high T$_c$ superconductors involve layered materials and magnetic fields applied on an arbitrary direction with respect to the layers. When the anisotropy is very large, Cooper pair currents can circulate either within or perpendicular to the layers. Thus, tilted magnetic fields lead to intertwined lattices of Josephson and Abrikosov vortices, with quantized circulation across and within layers, respectively. Transport in such intertwined lattices has been studied in detail, but direct observation and manipulation of vortices remains challenging. Here we present magnetic force microscopy experiments in tilted magnetic fields in the extremely quasi-two dimensional superconductor $Bi_{2}Sr_{2}CaCu_{2}O_{8}$. We trigger Abrikosov vortex motion in between Josephson vortices, and find that Josephson vortices in different layers can be brought on top of each other. Our measurements suggest that intertwined lattices in tilted magnetic fields can be intrinsically easy to manipulate thanks to the mutual interaction between Abrikosov and Josephson vortices.
\end{abstract}
\maketitle

\flushbottom

\section{Introduction}

Strongly anisotropic layered superconductors hold Josephson vortices across layers in parallel magnetic fields \cite{Koshelev1999}. Josephson vortices have no core, because the phase winding occurs in the non-superconducting region between the layers. When the magnetic field is applied perpendicular to the layers, Abrikosov vortices with a core nucleate within each layer. Abrikosov vortices in layered superconductors are composed by columns of disk-like pancake vortices and their shear modulus is significantly smaller than the shear modulus of Abrikosov vortices in isotropic superconductors. In the absence of Josephson vortices or pinning the pancake vortices form
straight lines along the $c$-axis. But in tilted magnetic fields, there is a finite in-plane magnetic field component and Josephson vortices appear. The mutual interaction between the columns of pancake vortices and the Josephson vortices leads to deformed combined lattices consisting of tilted vortex lines having kinks and different arrangements composed by layers with alternating pancake and Josephson vortices \cite{Grigorenko01,Curran2018}. In Ref.\cite{PhysRevLett.88.147002} it has been shown that the pancake vortices in such composite vortex lattices in tilted magnetic fields attract each other at large distances when located on top of a Josephson vortex. This long range attraction shapes the structure of combined lattices and leads to the formation of chains of pancake vortices along Josephson vortices. The chains decorate Josephson vortices, because pancake vortices sit rather close to each other along the lines of Josephson vortices. This occurs even at small perpendicular magnetic fields.

The vortex attraction in tilted magnetic fields is quite a general phenomenon and is also expected in superconductors that are not extremely anisotropic, such as YBa$_2$Cu$_3$O$_7$ and 2H-NbSe$_2$. It is caused by the current distribution produced by tilted magnetic fields when superconducting properties are anisotropic. Supercurrents then circulate on complex paths, which consist of ellipsoids whose shape depends on the tilt of the magnetic field with respect to the crystalline direction and this easily leads to the appearance of a minimum in the interaction potential between vortices \cite{Buzdin90,BuzdinSimonov91,Grishin90,Grishin92,PhysRevLett.69.2138,PhysRevLett.68.3343,PhysRevB.46.8425,PhysRevB.96.174516}.

\begin{figure*}[tb]
\includegraphics[scale=0.7]{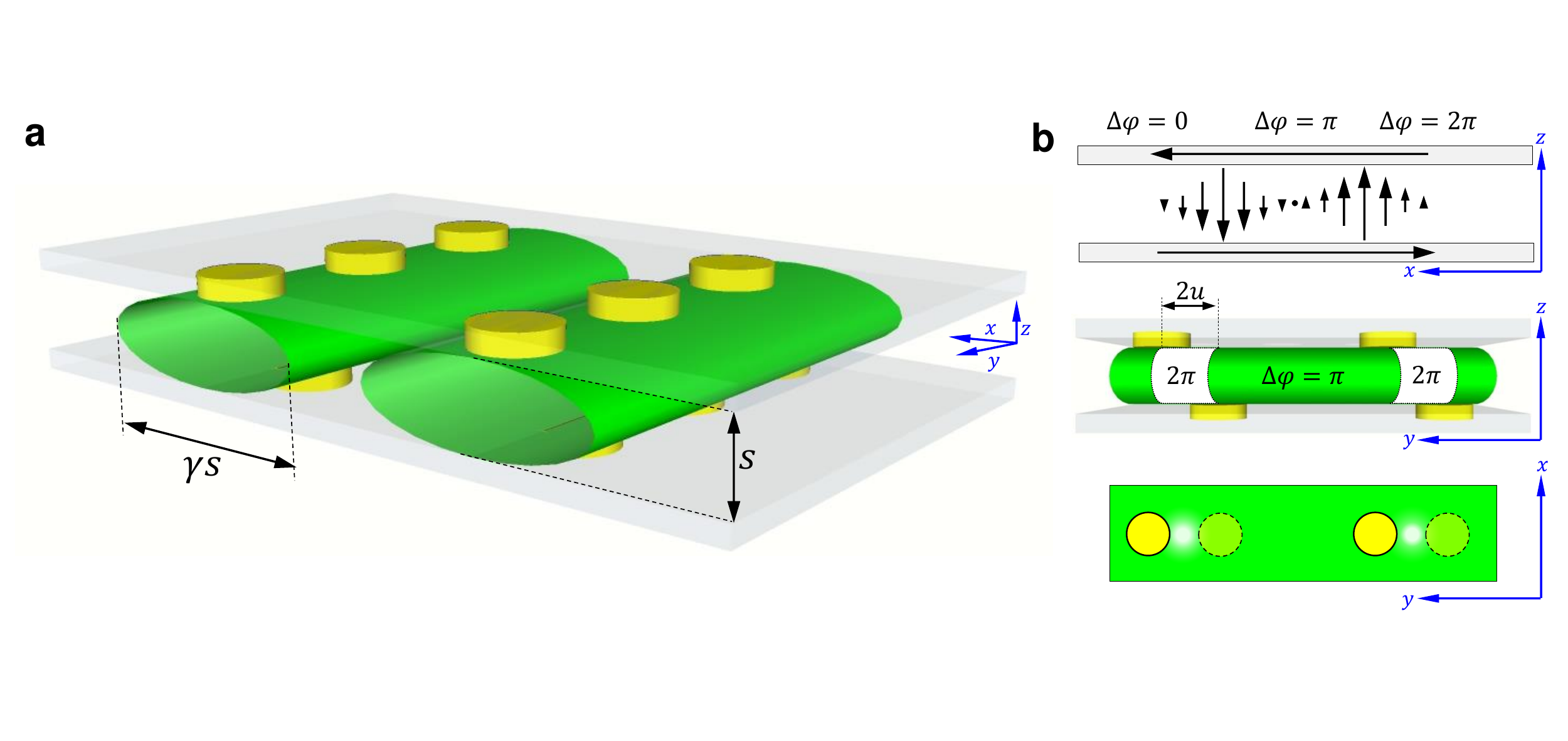}  
\vskip 0cm  
\caption{{\bf Crossing lattices of pancake and Josephson vortices.} In {\bf a} we show
superconducting layers by grey planes. Josephson vortices are represented by
tubes of elliptical cross-section with axis $\gamma s$ and $s$ where $\gamma$
is the anisotropy parameter and $s$ the interlayer distance. Yellow
disks represent pancake vortices, whose size in the schematic picture is of the order of the circulating current distribution, i.e. of order $\lambda_{ab}$. The pancake vortices are located along the Josephson vortex in the top and bottom planes. In the upper panel of {\bf b} we show
a lateral cross section through a Josephson vortex (at a position along $x$ free of pancake vortices). Josephson current is shown
by black arrows between layers. The phase difference $\Delta\varphi$ between layers as a
function of the position is also given. The current in each layer is shown by
black arrows within each layer. The lateral size of the Josephson vortex is
approximately of $\lambda_J=\gamma s$. In the middle panel we show the phase difference
between consecutive layers along a line through the center of a Josephson
vortex ($y=0$), taking into account the presence of pancake vortices (yellow
disks). In the white areas $\Delta\varphi= 2\pi=0$ and in the green areas $
\Delta\varphi= \pi$, because pancake vortices are displaced to each
other.  The currents circulating across a Josephson vortex lead to oppositely oriented
Lorentz forces in each layer. The displacement $u$ is induced by the Lorentz force and provides an energy gain thanks to the disappearance of the phase difference $\Delta\varphi=2\pi=0$ between pancake vortices. In the bottom panel we show a view of a Josephson vortex with pancake
vortices from the top. This is highly schematical, detailed calculations provide intricate patterns depending on pancake vortex density and relative sizes of pancake and Josephson vortices, see e.g. Refs.\protect\cite{PhysRevB.67.134501,PhysRevB.71.174507,Grigorenko01,Matsuda01,PhysRevB.66.014523,Koshelev1999,PhysRevLett.89.217003,Kirtley2010,0953-2048-27-6-063001,PhysRevB.46.366,0953-8984-17-35-R01}.}
\label{Fig1}
\end{figure*}

The phase diagrams with interwined Josephson and pancake lattices in highly anisotropic layered superconductors have been intensively studied as a function of the magnetic field and temperature \cite{PhysRevB.67.134501,PhysRevB.71.174507,Grigorenko01,Matsuda01,PhysRevB.66.014523,Koshelev1999,PhysRevLett.89.217003,Kirtley2010,0953-2048-27-6-063001,PhysRevB.46.366,0953-8984-17-35-R01,Zeldov2011}. Josephson vortices have circulating currents on an area defined by the interlayer separation $s$ and the in-plane length $\lambda_{J}$ (Fig.\ref{Fig1} {\bf a} and upper panel of {\bf b}). Pancake vortices have currents circulating on a disk of size of order of the in-plane penetration depth $\lambda_{ab}$. Pancake vortices lie in each layer. Thus, there are pancake vortices in the two layers forming the Josephson vortex. The mutual interaction between the two kinds of vortices leads to a shift of the pancake vortices lying in each layer along the Josephson vortex. Thus, pancake vortices are not exactly on top of each other, as usual in an Abrikosov lattice in perpendicular fields, but are slightly displaced along the length of the Josephson vortex. This is due to the Lorentz force from the intralayer currents, which are directed opposite to each other in each layer. There is a phase change of $2\pi$ across each vortex, once across layers for a Josephson vortex and once within each layer for a pancake vortex. But when pancake vortices are not exactly on top of each other in two adjacent layers, the $\pi$ phase slip across the two layers vanishes in between displaced pancake vortices. Thus there is no Josephson vortex in between two pancake vortices at two different locations on two different layers (Fig.\ref{Fig1} {\bf b} middle and bottom panels). This results in an energy gain, which is of order of the Josephson coupling energy of the section of the Josephson vortex that disappears in between pancake vortices. When repeating along the whole sample, the presence of pancake vortices on top of the Josephson vortex can thus significantly decrease the overall energy cost of tilted magnetic fields. The interaction between neighboring pancake vortices in the same layer provides a balance and there is an equilibrium distance for pancake vortices along the Josephson vortex. The variation of this distance with tilt and value of the magnetic field and with temperature leads to the equilibrium phase diagrams discussed in literature \cite{Koshelev1999,Grigorenko01,0953-8984-17-35-R01}.

Here we study slightly underdoped $Bi_{2}Sr_{2}CaCu_{2}O_{8}$ (BSCCO), which is a highly anisotropic
superconductor consisting in $CuO_2$ bilayers separated by a distance $s=1.5$ $nm$,
with $Ca$, $SrO$ and $BiO$ layers in between. We find an anisotropy factor $\gamma\approx 1000$ ($\gamma=\lambda_{c}/\lambda_{ab}=\xi_{ab}/\xi_{c}$ with $\lambda_{ab}$ and $\xi
_{ab}$ the in plane and $\lambda_{c}$ and $\xi_{c}$ the out of plane penetration depths and coherence lengths respectively). We mostly study Josephson vortices with an in-plane magnetic field. We decorate the Josephson vortices by pancake vortices and work in a strongly out of equilibrium situation. We obtain vortex patterns by modifying the amount of pancake vortices on top of the Josephson vortices through repeated heating and cooling. We show that, under such out of equilibrium situation, pancake vortices can be displaced in between Josephson vortices and that Josephson vortices can be manipulated by changing the in-plane direction of the magnetic field.

\begin{figure*}[ptb]
\includegraphics[width=0.95\textwidth]{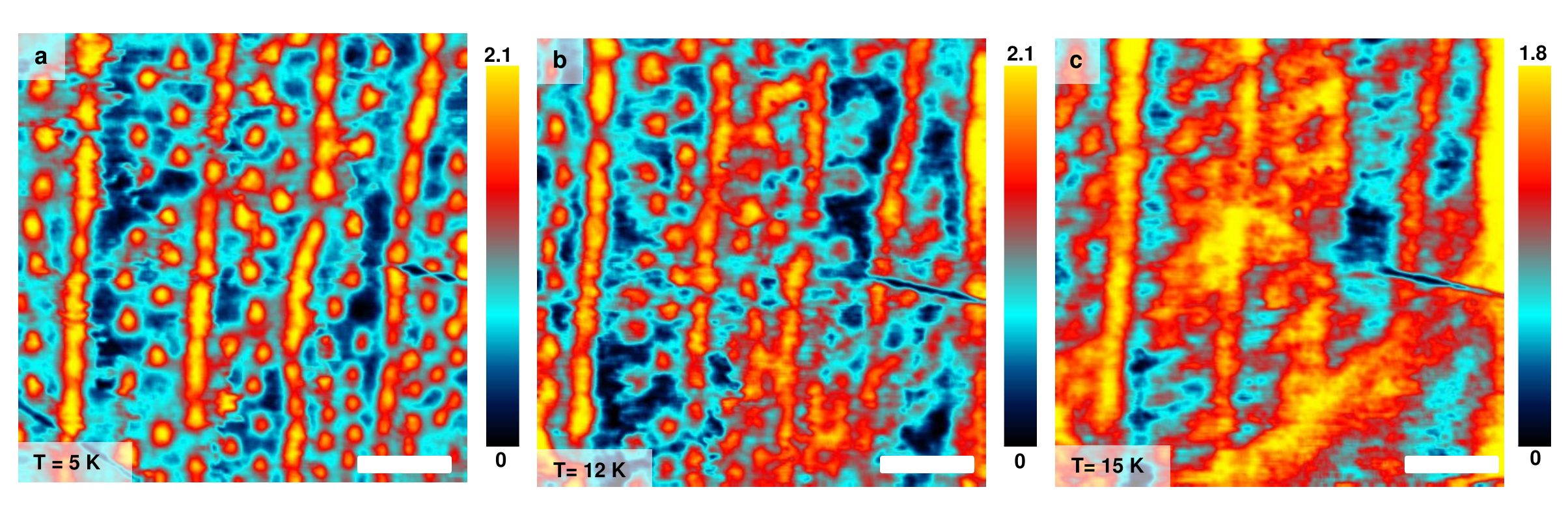}  
\vskip 0cm  
\caption{{\bf Tip induced vortex motion far below the depinning temperature.} 
Magnetic Force Microscope (MFM) images taken at 5 K {\bf a}, 12 K {\bf b} and 15 K {\bf c} at a magnetic field of 200 G
applied along the y-axis of the images. Prior to the application of the parallel magnetic field, we have induced pancake vortices to decorate Josephson vortices, as described in the text and the Supplementary Information. The field of view is the same for all
images. The contrast is given by the color bars on the right in changes in degrees of the phase of the cantilever oscillation (see Methods section) and white bars in the images are of $2.4$ $\mu m$ size.}
\label{Fig2}
\end{figure*}

\section{Results}

\subsection{Interaction between the MFM tip and pancake vortices}

We start by discussing briefly the decoration of Josephson vortices with pancake vortices and the invasive action of the MFM tip. In Fig.\,\ref{Fig2} {\bf a} we show a Magnetic Force Microscope (MFM) image of a crossing lattice of Josephson and pancake vortices obtained at 5 K. To obtain this image we first apply a magnetic field of 50 G perpendicular to the layers. This introduces pancake vortices in the sample. We then apply 200 G parallel to the layers and ramp the perpendicular component down to zero. We briefly heat the sample to above 50 K and cool down again. This leads to a re-arrangement of pancake vortices that position themselves along the Josephson vortices, creating the vertical rows shown in Fig.\,\ref{Fig2} {\bf a}. 

By heating above 50 K again, we can reduce the amount of pancakes in between Josephson vortices until we reach a non-equilibrium situation with an in-plane magnetic field and pancake vortices lying on top of Josephson vortices, discussed below.

It is quite convenient to discuss the invasive action of the MFM tip using the location shown in  Fig.\,\ref{Fig2} {\bf a}, which was obtained before repeated heating and has quite a large amount of interstitial vortices. When we increase the temperature from 5 K to 15 K (Fig.\,\ref{Fig2} {\bf b,c}), we observe that the size of pancake vortices in the images is considerably increased when reaching 15 K (see also Supplementary Information). At higher temperatures, we do not resolve pancake vortices anymore. This increase in apparent size is due to thermal motion. Thermal motion is activated by the wiggling action produced by the MFM tip when scanning on pancake vortices, in a similar way that vortex lattice melting is favored by a dithering AC magnetic field. To complete a square image, the tip scans along a line (say, the x-axis, as in Fig.\,\ref{Fig2}) and then moves to the next. Thus, it moves fast along one direction (say, x-axis) and slow along the other direction (say, y-axis). The tip scans several times over each pancake vortex when moving along the fast scan direction. As shown in Ref.\cite{Avraham2001}, a dithering magnetic field considerably reduces vortex pinning. It has been shown previously that scanning a magnetic tip produces such a dithering field locally and enough vortex shaking to eliminate pinning \cite{Straver2008,Auslaender2008,doi:10.1142/S0217979210056384,PhysRevB.80.054513,Embon2015,PhysRevB.80.054513,PhysRevE.87.022308,doi:10.1021/acs.nanolett.5b04444}. This allows to manipulate pancake vortices thanks to the dithering magnetic field produced by the tip motion during scanning. It is interesting to note that the temperature where we observe depinning, which is of 15 K, is far below the depinning temperature of BSCCO (a few K below T$_{c}=$ 88 K) \cite{Schwarz2010,Koshelev1999}. The depinning temperature was reduced down to about 30 K in a similar sample by dithering magnetic fields in Ref.\cite{Avraham2001}. In our experiment, we reach temperatures about half that value, suggesting that the spatial gradient of the magnetic field plays a role and increases the force on vortices during this process. By increasing and decreasing the temperature in a relatively small interval, we can thus trigger pancake vortex motion using our MFM tip.

\begin{figure*}[ptb]
\includegraphics[width=0.95\textwidth]{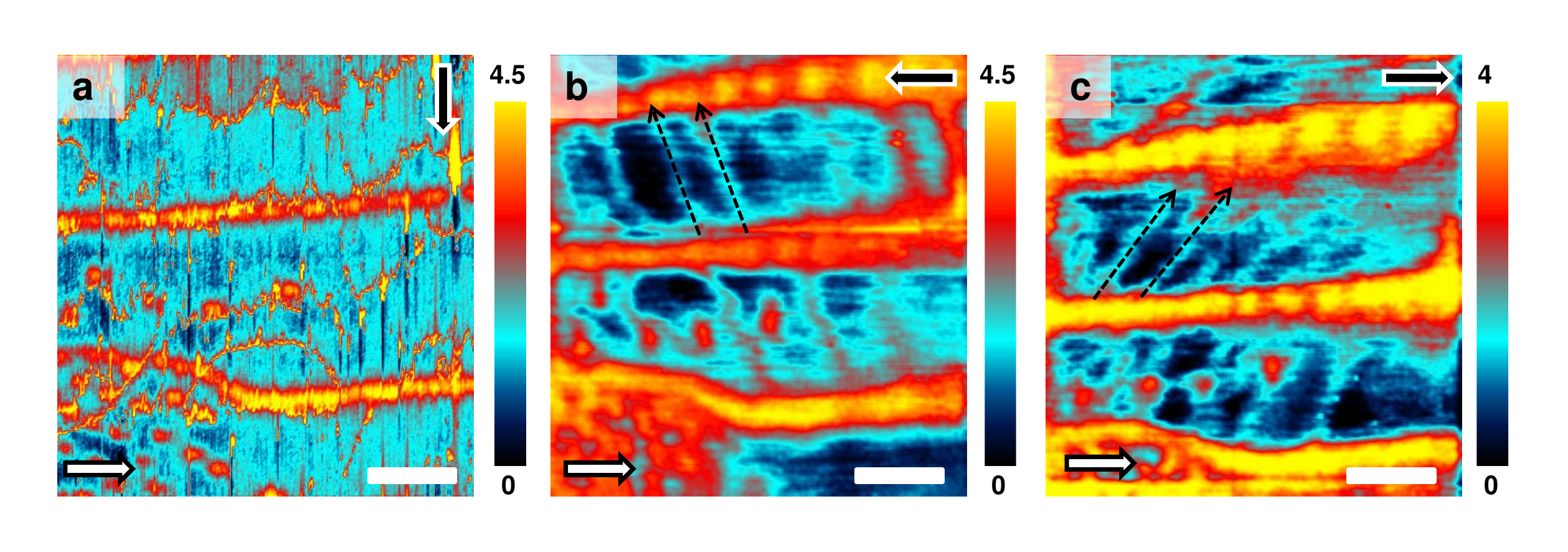}  
\vskip 0cm  
\caption{{\bf Pancake vortex motion across Josephson vortices.} MFM images at a magnetic field of 200 G parallel to the layers. The in-plane direction of the magnetic field is shown by the white arrows. The scanning procedure is detailed in the Methods section. The direction at which the tip scans fast is given by the black arrows. Image {\bf a} is taken at 5.5 K  and {\bf b,c} at 12 K. White bars in the images are of $2.6$ $\mu m$ size. In {\bf b,c} we observe pancake vortex motion along the slow scanning direction (towards the top of the image) across Josephson vortices. This provides traces of pancake vortices that are slightly to the left in {\bf b} and slightly to the right in {\bf c}. We mark two of these traces by black dashed lines with arrows. The contrast is given by the color bars on the right in changes in degree of the phase of the cantilever oscillation (see Methods section).}
\label{Fig3}
\end{figure*}

\subsection{Moving pancake vortices between Josephson vortices}

We now study a situation where we have built a decorated Josephson vortex lattice with only a few interstitial pancake vortices and show that we can move pancake vortices across Josephson vortices using the MFM tip. In the Fig.\,\ref{Fig3} {\bf a}, the tilt is along the x axis of the image and most
pancake vortices are aligned and pinned at the Josephson vortices. The in plane
component of the magnetic field in this image is aligned with the slow axis of
the scan direction. We then increase the temperature to 12 K and exchange slow and fast scanning directions. We observe the images shown in Figs.\,\ref{Fig3} {\bf b,c}. Pancake vortices move across Josephson
vortices. Note that vortex motion is perpendicular to the slow scan direction. This corresponds to the view that the dithering AC magnetic field reduces or eliminates vortex pinning. The fast scanning acts to reduce the pinning and the vortices are slowly pushed along the slow scan direction when the tip moves from one line to the next. Correspondingly, vortex motion is not exactly perpendicular to the Josephson vortices, but with a small angle (see also Supplementary Figure S3). The angle depends on the direction of the fast
scan, from right to left (Fig.\,\ref{Fig3} {\bf b}) or viceversa  (Fig.\,\ref{Fig3} {\bf c}).

\subsection{Approaching and crossing Josephson vortices}

\begin{figure*}[ptb]
\centering
\includegraphics[width=0.95\textwidth]{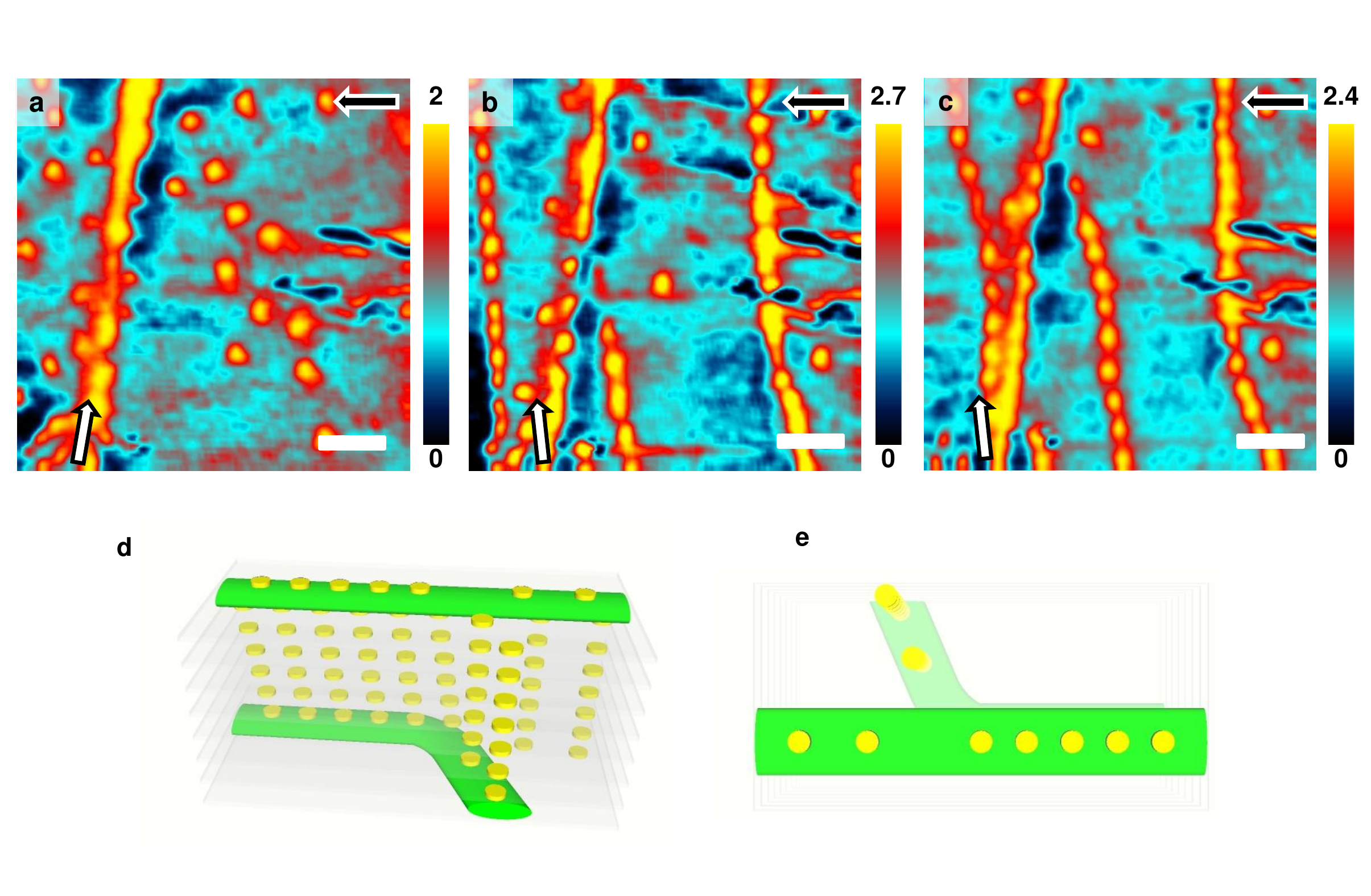}  
\vskip 0cm  
\caption{{\bf Attraction between crossing Josephson vortices.} In {\bf a} we apply a magnetic field of 200 G tilted 5 degrees with respect to the y-axis of the image (white arrow). In {\bf b} we show the same field of view after modifying the in-plane tilt of the magnetic field by 10$^{\circ}$ (white arrow) and heating once up to 20 K. In {\bf c} we show the same image, after heating anew to 20 K. Images are taken at 5.3 K. Fast scanning direction is given by the black arrow. The contrast is given by the color bars on the right in changes in degree of the phase of the cantilever oscillation (see Methods section) and white bars in the images are of $2.6 \mu m$ size. In {\bf d} we show a schematical view of Josephson vortices (green tubes), layers (grey planes) and pancake vortices (yellow disks). Josephson vortices cross in different layers. When they join, they share the same pancake vortex columns (view from the top, {\bf e}), leading to a decrease in energy that produces an effective attractive interaction between Josephson vortices decorated by pancake vortices.}
\label{Fig4}
\end{figure*}

We now produce crossing Josephson vortices using the combined action of pinning and changing the direction of the in-plane magnetic field. We start by nucleating a single decorated Josephson vortex in a field of view that has a linear defect. The linear defect is actually a large circular wrinkle, shown in the Supplementary Figure 8. We apply the in-plane magnetic field exactly along the defect and observe that we nucleate a Josephson vortex along the defect (Fig.\,\ref{Fig4} {\bf a}). We then turn the in-plane magnetic field by about 10 degrees and heat and cool several times. We observe further Josephson vortices in the field of view. These are Josephson vortices that are oriented along the tilt of the magnetic field and are thus at an angle with respect to the Josephson vortex pinned at the wrinkle (Fig.\,\ref{Fig4} {\bf b,c}). Heating and cooling from Fig.\,\ref{Fig4} {\bf b} to Fig.\,\ref{Fig4} {\bf c} leads to motion of the decorated Josephson vortex lines. As we can see in the left part of the Fig.\,\ref{Fig4} {\bf c}, the Josephson vortices do not repel. The leftmost Josephson vortex has approached the pinned Josephson vortex and the Josephson vortex in the middle of the image Fig.\,\ref{Fig4} {\bf c} results from the motion of a Josephson vortex from the lower left corner to the middle of the image. In this process, the latter vortex has joined to pancakes that were previously located in between Josephson vortices in the middle of the image in Fig.\,\ref{Fig4} {\bf b}. The observed lines of Josephson vortices, lying in different layers (as discussed below), clearly cross each other. The density of pancake vortices along the crossing Josephson vortices changes when they are close to each other. For instance, the upper part of the leftmost Josephson vortex in Fig.\,\ref{Fig4} {\bf c} has a smaller density than the lower part.

\section{Discussion}

Most previous experiments dealing with crossing Josephson and pancake vortex lattices were made close to liquid
nitrogen temperatures using non-invasive imaging such as scanning Hall microscopy. Modifications in the Josephson vortex structure can then be made by changing the direction of the magnetic field \cite{Grigorenko01,0953-8984-17-35-R01}.

In our experiments, we take advantage of dithering field produced by the moving MFM tip. The magnetic field due to the scanning MFM tip oscillates with a frequency of the order of several tens or hundreds of Hz (see Methods section for details). Previous experiments used the same effect to move vortices in other cuprates with less in-plane vs out-of-plane anisotropy and in thin films of low temperature superconductors \cite{Straver2008,Auslaender2008,doi:10.1142/S0217979210056384,PhysRevB.80.054513}. Macroscopic experiments have shown vortex shaking at frequencies ranging from tens of Hz to hundreds of kHz also in both low temperature and in cuprate superconductors, including BSCCO \cite{PhysRevB.84.012508,PhysRevE.87.022308,Avraham2001}. This provides the opportunity to manipulate vortices and show their vortex motion and mutual interaction at low temperatures, even in presence of strong pinning and interacting Josephson and pancake vortex lattices in BSCCO. 

When the magnetic field is parallel to the layers, say of $200$ $G$, the distance between Josephson vortices along the $c$-axis $a_{z}\approx 10$ nm, which is nearly an order of magnitude larger the distance $s$ between superconducting CuO$_{2}$ bilayers ($s=$ 1.5 nm). The pancakes decorate  two
adjacent rows of Josephson vortices. These occupy different interlayer spaces, shifted by $\ a_{z}/2\approx 6$ nm \cite{PhysRevB.68.094520}. Therefore, when we drag the row of pancake vortices toward
the pancake vortices decorating another Josephson vortex in the experiment shown in (Fig.\,\ref{Fig4} {\bf b,c}), the approach and crossing occurs between different layers. When two Josephson vortices cross (Fig.\ref{Fig4}) the rows of pancakes are additionally distorted by the flux from the pancakes in the Josephson vortex immediately below (Fig.\,\ref{Fig4} {\bf c,d}). The resulting decrease of the vortex energy provides a mechanism of a short range attraction between rows of Josephson vortices.

The energy per unit area of the Josephson interaction between superconducting layers is given by $E_J(1-cos\Delta\varphi)$, where $E_J=\frac{s\Phi_0^2}{16\pi^3\lambda_{ab}^2\lambda_J^2}$ and $\Delta\varphi$ is the phase difference between layers \cite{PhysRevB.68.094520}. By contrast, the energy due to the mutual attraction between two pancakes to lie on top of each other is $E_M=u^2\Phi_0^2(\frac{1}{32\pi^2\lambda_{ab}^4}ln(\frac{\lambda_{ab}}{u}))$ where $u$ is the displacement of pancake vortices between layers (Fig.\ref{Fig1} {\bf b}). When pancake vortices are displaced along the Josephson vortex, the total energy is of order of $E_J\lambda_J u+E_M$. The crossing of the pancake stack with a Josephson vortex produces stack’s deformation and leads to the energy decrease $\delta E\approx -E_J\lambda_{ab}^2 \approx -\frac{\Phi_0^2}{\gamma^2 s}$ \cite{PhysRevB.68.094520}.

When two Josephson vortices are one under the other, the same pancake stacks cross them, which doubles the energy gain due to the crossing. This leads to the attraction between these Josephson vortices. The corresponding attractive force (per one stack crossing) can be estimated as $F_{att}\approx \frac{\delta E}{\lambda_J} \approx \frac{\Phi_0^2}{\gamma^3 s^2}$, where we took into account that the effective mutualisation of the pancake stacks occurs when the distance between Josephson vortices is less than $\lambda_J$. The equilibrium distance between pancake stacks decorating Josephson vortex is $\approx 4\lambda_{ab}$ and the attraction force $F_{att}$ (per unit length) between Josephson vortices can be estimated as $F_{att} \approx \frac{\Phi_0^2}{\gamma^3 s^2 \lambda_{ab}} \approx 5$ $\times$ $10^{-6}$ Nm$^{-1}$. This force is comparable to the repulsive force $F_{rep}$ between Josephson vortices $F_{rep} \approx \frac{\Phi_0 H_{c1}}{\lambda_{ab}} \approx \frac{\Phi_0^2}{\gamma \lambda_{ab}^3}$. This makes it possible to obtain the metastable configuration of Josephson vortices with crossing and branching as observed in the experiments and shows that Josephson vortices lying in different planes have a tendency to share pancakes.

Ref.\cite{PhysRevLett.89.217003} has extensively discussed the interaction between linear defects in the CuO$_2$ bilayers and crossing lattices using Scanning Hall Microscopy and tilted magnetic fields at liquid nitrogen temperatures. It has been shown that pancake vortices can be preferentially pinned at linear defects, favoring that Josephson vortices are located along these lines. Usually, linear defects lie perpendicular to each other and are of a finite length. This can lead to kinks in the pancake vortex distribution when abruptly modifying the in-plane magnetic field, because some vortex sections remain pinned whereas others free themselves. Images then show Josephson vortices that fork, similar to those we see in our experiment. Experiments in \cite{PhysRevLett.89.217003} were made close to liquid nitrogen temperatures, whereas we work at much lower temperatures. Furthermore, they abruptly modified the magnetic field, whereas we use a superconducting coil which is slow. Finally, the phenomenon of forking appears here in a strong linear defect. This is a true wrinkle that leads to a considerable modification of the surface, suggesting that the strongly pinned Josephson vortex is at the surface. Linear defects might be related to small steps at the surface of finite size. We did not manage to pin and cross vortices with such defects, in spite of trying. This shows that the temperature is playing a relevant role in vortex manipulation. Manipulating vortices at low temperature requires very well defined pinning centers. Nevertheless, it is clear that our results are also influenced by quenched disorder. For instance, our images show pancake vortex lines that are not completely straight and there are often pancake vortices in between lines.

We can now discuss the observed drag of pancake vortices across Josephson vortices by the tip (Fig.\ref{Fig3}). The force needed to move pancakes can be of order of $F_{att}$, which gives about 5 pN if we consider a length of the pancake column of order of a micron. This suggests that it is relatively simple to move pancakes across the Josephson vortices, as made in Fig.\ref{Fig3} {\bf b,c}. Note that in that experiment, the temperature is close to the depinning threshold, which explains the decrease in the depinning force with respect to experiments made at lower temperatures. Furthermore, the pancake vortices can be also pushed into the neighboring Josephson vortices (arrows in Fig.\ref{Fig3} {\bf b,c}), showing again that the effective attractive interaction of pancake vortices inside a Josephson vortex strongly influences vortex dynamics.

It is also worth to discuss the lateral motion of pancake vortices in between Josephson vortices (dashed lines in Fig.\ref{Fig3} {\bf b,c}). The crystal is aligned in such a way that the nodes of the d-wave superconducting order parameter are located along the diagonals in Fig.\ref{Fig3} {\bf a-c}. Macroscopic critical current experiments in the flux flow regime show indications for enhanced mobility along these directions \cite{PhysRevLett.97.067003}. It thus is tempting to think that the additional quasiparticles present along nodes favor, together with the lateral push of the tip, the direction of motion in these experiments. 

Vortex entanglement, which is a usual situation in superfluids, is more difficult to find experimentally in superconductors \cite{RevModPhys.66.1125,PhysRevLett.60.1973}. Using manipulation with MFM, it is in principle possible to entangle vortices in layered superconductors. The entangled state consists of interacting groups of vortices, forming, for instance, helices or disordered tangles and is different from the situation where vortices cut each other \cite{PhysRevLett.92.157002,PhysRevLett.75.1380}. The entangled state results in strongly reduced vortex mobility due to the interaction among vortices, whereas cutting increases vortex mobility and reduces pinning \cite{PhysRevB.42.9938,PhysRevLett.67.3176,PhysRevB.50.10294,PhysRevLett.80.1070}. The low coupling between pancake vortices lying in different planes makes it difficult to produce vortex entanglement (instead of flux cutting) in layered two-dimensional systems as BSCCO \cite{PhysRevLett.92.157002}. Previous pancake vortex manipulation experiments using a MFM in YBa$_2$Cu$_3$O$_{6.4}$ show that, even in perpendicular fields, it is possible to produce kinks in lines of pancake vortices \cite{PhysRevB.79.214530}. Here we have shown that, in parallel magnetic fields, Josephson vortices lying in different planes can move to be one on top of the other thanks to the attractive interaction among pancake vortices. The modifications in the interaction between vortices produced by tilted magnetic fields have been previously considered and depend on the anisotropy of the superconducting properties \cite{PhysRevB.43.10482,PhysRevLett.67.3176}. However, the strongly non-equilibrium situation we unveil here in tilted magnetic fields has not been treated until now and significantly adds to the previous observation of kinks in Abrikosov pancake vortex lines \cite{PhysRevB.79.214530}. In particular, the effective attraction between Josephson vortices that we unveil here shows that crossing lattices of Josephson and pancake vortices can have strong interactions among them. If this leads to increased pinning and avoids flux cutting is not fully clear from our data. We also see that crossing Josephson vortices are quite stable and remain when rotating the magnetic field (see Supplementary Information). This suggests that pinning instead of flux cutting is indeed dominating the behavior at very low magnetic fields.

In summary, we have shown how to manipulate pancake and Josephson vortices by combining a MFM tip and rotating magnetic fields well below the critical temperature. We find that pancake vortices can be dragged in between Josephson vortices and unveil an effective attractive interaction between pancake vortices lying inside Josephson vortices.

\section{acknowledgments}

Work done in Madrid was supported through grant numbers FIS2017-84330-R, MDM-2014-0377, MAT2014-52405-C2-2-R,
RYC-2014-16626 and RYC-2014-15093 of AEI, by the Comunidad de Madrid through program
Nanofrontmag-CM (S2013/MIT-2850), by the European Research Council PNICTEYES
grant agreement no. 679080, by EU Flagship Graphene Core1 under Grant
Agreement no. 696656, by French ANR project OPTOFLUXTRONICS and by EU COST CA16218 Nanocohybri. SEGAINVEX at UAM is
also acknowledged. We also acknowledge the support of Departamento Administrativo de Ciencia, Tecnolog\'ia e Innovaci\'on, COLCIENCIAS (Colombia) Convocatoria 784-2017 and the Cluster de investigaci\'on en ciencias y tecnolog\'ias convergentes de la Universidad Central (Colombia).

Correspondence should be sent to hermann.suderow@uam.es.

\section{Methods}

\subsection{Sample and MFM setup}

We have measured a BSCCO single crystal with onset T$_{c}$ of 88 K, i.e. on the slightly underdoped regime \cite{PhysRevLett.86.886,PhysRevB.66.132505}. We mounted a sample in the sample holder of a low temperature MFM and inserted the microscope in a three axis vector
magnet. The set-up is described in Ref.\,\cite{Galvis2015}. The crystal is a millimeter sized rectangular plate with a thickness of about 0.3 mm. We cleaved the crystal and aligned it with the main axis of our coil system, to apply the z-component of the
magnetic field along the c-axis and align x and y components of the magnetic
field with the in-plane crystalline axis and the x and y directions of the scanning field of view. We prepared our tip by leaving a coercive magnetic field on the tip that is
parallel to the z component of the applied magnetic field. We always applied
magnetic fields well below the switching or saturating fields of the
tip \cite{Galvis2015}.

\subsection{MFM images}
The color scale of the images provides the change in the phase of the oscillation of the cantilever with respect to the excitation as a function of the position. The interaction between tip and sample can be influenced by the local magnetic field, but also by a variety of other interactions, such as electrostatic interactions, stray magnetic fields coming from close-by areas of the sample or small changes in the magnetic properties of the tip. Thus, the color scale cannot be easily transformed into a magnetic field. When the tip is far enough, however, magnetic force is expected to dominate other effects. Thus, we can consider, given the mentioned caveats, that, in each image, the color scale provides relative measure of the changes in the magnetic field as a function of the position. As a matter of fact, it is quite clear from the images that, when vortices wiggle and their apparent size increases, the corresponding contrast  in the MFM images (difference in phase across the whole image) is also reduced. Furthermore, tests such as the density of pancake vortices with respect to the tilt of the magnetic field (see Supplementary Figure 7), provide results in good agreement with expectations from simple geometrical constructions giving the z-component of the applied magnetic field. Images are rendered using WSXM software \cite{doi:10.1063/1.2432410} and additional software developed by some of the authors that will be described in another publication.

\subsection{Scanning process and dithering field}

The scanning process is optimized to make fast images. We make lines by going forth and back with the tip, and then advance one further step in the other direction. The direction for back and forth scanning is the fast scanning direction, and the direction perpendicular to it is the slow scanning direction. We scan close to the sample to record topographic information in AFM (Atomic Force Microscopy) mode when we go forth and retrace the tip about 50 nm and scan back the same line to record the magnetic image (MFM mode), or viceversa. The left to right and right to left motion of vortices in Fig\,3{bf b} and {\bf c} is produced by the scan close to the sample in AFM mode. The scanning speed is of about a second per scanning line and we usually make images of 256$\times$256 points. The tip needs between between 0.01 and 0.05 seconds to scan across a vortex, which gives an oscillating magnetic field of between several tens and a hundred Hz.

\subsection{Calculations of distances between vortices and sample parameters}

Following Ref.\cite{PhysRevLett.88.147002}, we can calculate the equilibrium distance between pancake vortices and find $d\sim4\lambda_{ab}\sim1\mu m$ for $B_{||}=200$ $G$.  Using the distance between the pancake rows $d_{rows}$ and the value of the magnetic field $B$ we can calculate the anisotropy factor of our BSCCO crystals $\gamma$ and the distance between Josephson vortices along the c-axis, $a_{z}$\cite{Koshelev2013,Koshelev1999,PhysRevB.38.2439}. We obtain $\gamma=2d_{rows}^2B/(\sqrt{3}\Phi_0)\approx 1000$ and $a_{z}\approx 10$ nm. The lateral size of the Josephson vortex is given by $\gamma s$ where $s$ is the distance between
CuO$_{2}$ bilayers and we find approximately a $\mu$m ($s$ being 1.5 nm).   

\subsection{Data availability}

The datasets generated during and/or analysed during the current study are available in the OSF repository, DOI 10.17605/OSF.IO/S3NCA.

\section{Author contributions}

AC and CM performed the experiments and analyzed the data, together with IG, EH, MG and FM. BSCCO samples were grown by TY and TK under the supervision of KK. Theoretical estimations and inputs for the data analysis were provided by AB. Manuscript was written by HS and IG with input from all authors, particularly by CM and EH. HS conceived the experiment and supervised the project together with CM.

\section{Competing interests}
The authors declare no competing interests.

\onecolumngrid
\newpage

\appendix

\section{Appendix I. Decorating a Josephson vortex lattice}

When we apply a magnetic field exactly in-plane, without any out of plane
component immediately after cooling down the first time, we often do not observe anything in the MFM images. To observe Josephson vortices, we prepare a specific decoration pattern by applying magnetic fields perpendicular to the layers, removing these and applying in-plane magnetic fields, to finally heat and cool until all pancakes are gone, except those lying on top of Josephson vortices.

In the Supplementary Figure\,\ref{Fig5} {\bf a} we show a lattice of pancake vortices obtained after applying a perpendicular magnetic field, removing it at low temperatures, applying an in plane magnetic field and increasing the temperature to 50 K. Increasing the temperature to higher values (60 K {\bf b}, 70 K {\bf c}) is usually enough to allow pancake vortices to leave the sample. Then, only those on top of the Josephon vortices remain.

\begin{figure*}[h]
\includegraphics[width=0.95\textwidth]{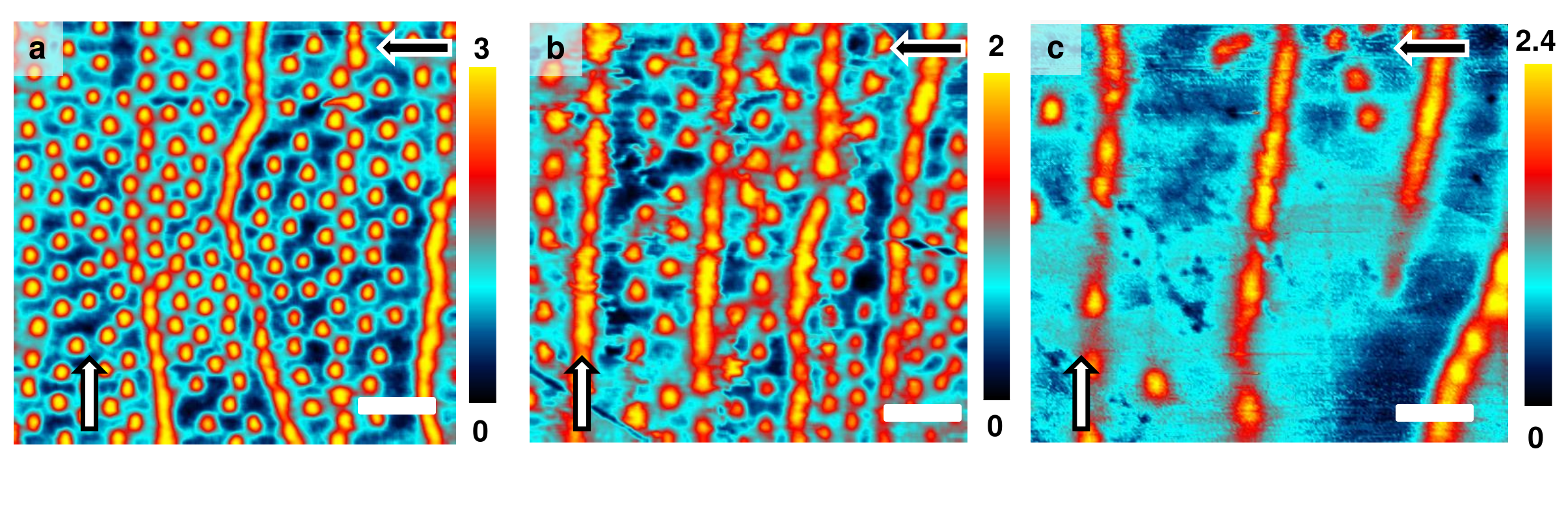}  
\vskip 0cm  
\caption{{\bf Decorating the Josephson vortex lattice.} MFM images of Josephson vortices decorated with pancake vortices. We first applied a magnetic field of 50 G along z, decreased it at 2K to
zero and applied an in-plane magnetic field of 200 G. We then increased the temperature to 50 K {\bf a}, 60 K {\bf b} and 70 K {\bf c}. Color scale gives changes in the phase of the oscillation of the cantilever and is related to the local magnetic field and a variety of other effects that have been minimized as far as possible, as described in the Methods section and as is usual in MFM experiments. White bars in the images are of $3 \mu m$ in size. Scanning direction is shown by the black arrow and the direction of the in plane magnetic field by the white arrow.}
\label{Fig5}
\end{figure*}

\section{Appendix II. Schematics of vortex motion induced by the magnetic tip}

In the Supplementary Figure\,\ref{Fig6} we schematically discuss vortex motion produced by scanning the magnetic tip. The motion of the tip along the fast scan axis Supplementary Figure\,\ref{Fig6} shakes vortices out of their pinning potential, by inducing an oscillatory magnetic field variation with a frequency above the dynamic depinning transition. Then, the tip pushes vortices with a force $F_{ts}$ in Supplementary Figure\,\ref{Fig6}. The tip acts on the vortices along the slow scan axis with a force $F_{tf}$.

In Supplementary Figure\,\ref{Fig7} we schematically describe motion of pancake vortices decorating
Josephson vortices. The motion is here just produced by the magnetic tip. The
motion acquires a lateral drift that depends on the direction of the scan. In particular, on the side on which the tip first interacts with the vortex.

\begin{figure*}[h]
\centering
\includegraphics[width=0.95\textwidth]{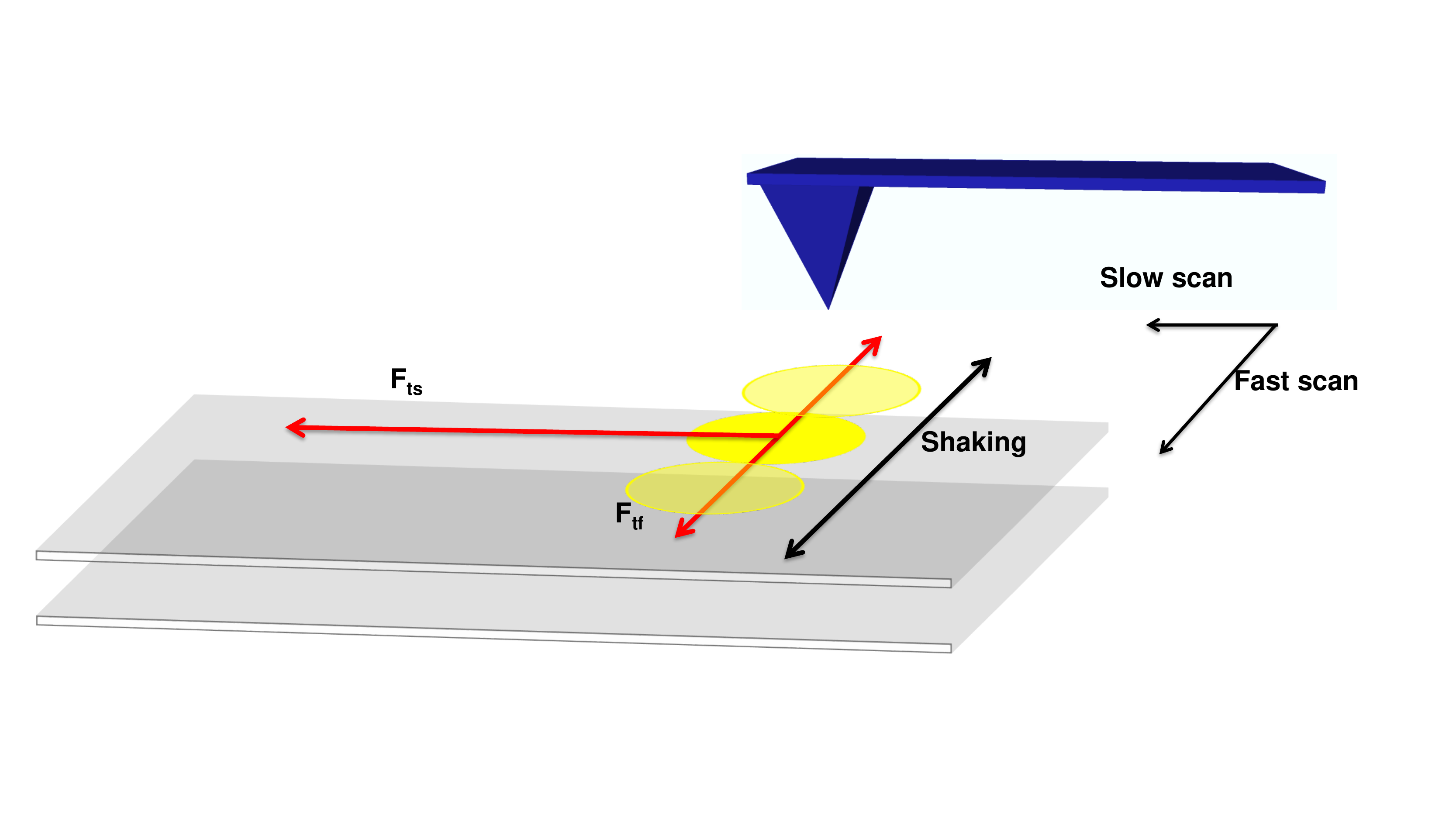}  
\vskip 0cm  
\caption{{\bf Schematics of vortex motion induced by the magnetic tip.} We represent the
tip and cantilever in blue and the forces from the cantilever on the vortices by red arrows. In {\bf a} we show the action of
the tip on pancake vortices, represented as yellow disks. Grey planes represent the superconducting layers. Red arrows are $F_{ts}$
and $F_{tf}$, the forces of the tip along the slow and fast scan axis
respectively.}
\label{Fig6}
\end{figure*}

\begin{figure*}[h]
\centering
\includegraphics[width=0.95\textwidth]{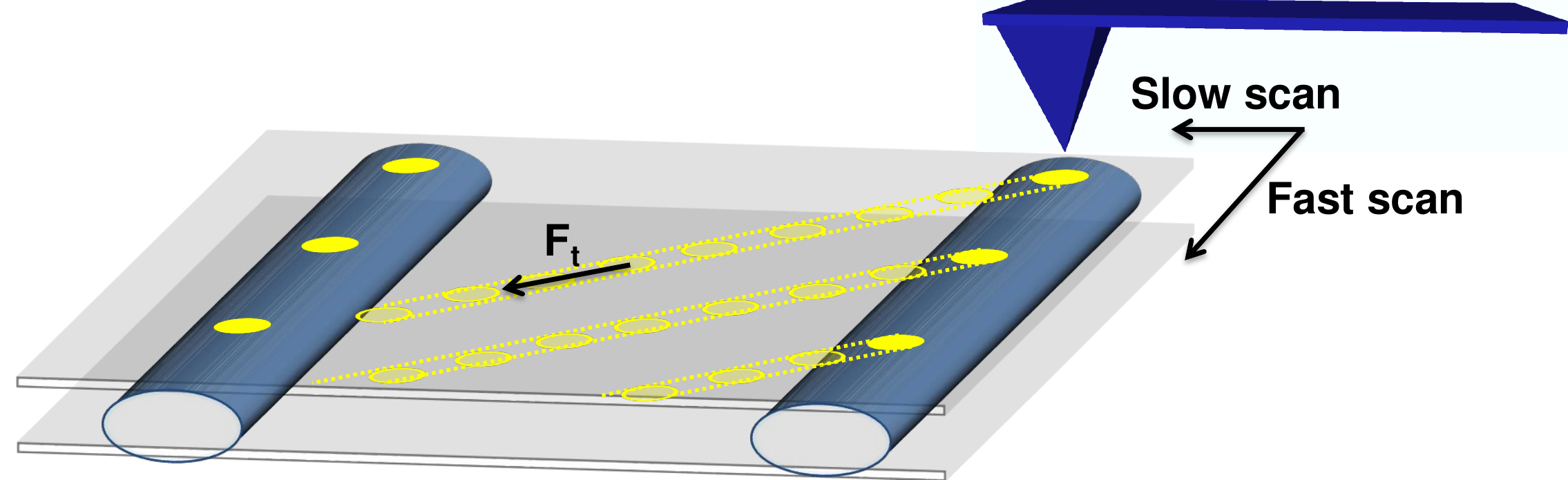}  
\vskip 0cm  
\caption{{\bf Schematics of motion of pancake vortices in between Josephson vortices induced by the magnetic tip.} We represent
the tip and cantilever in blue. Scanning directions are given by black arrows. Josephson vortices are represented by blue tubes with elliptical cross section. Superconducting layers are represented by grey plates. Pancake vortices are represented by yellow disks. The
MFM exerts here a force $F_{t}$ on the pancake vortices, with a lateral component parallel to the direction in which the tip moves during scanning.}
\label{Fig7}
\end{figure*}

\section{Appendix III. Changing the tilt of the magnetic field}

In Supplementary Figure\,\ref{Fig9} we show the result of modifying the direction of the in-plane magnetic field in an area where we have brought two Josephson vortices to cross each other. The experiments are made at low temperatures (about 5 K) and the magnetic field is of 200 G. When we heat up to 12 K we observe vortex wiggling, but the arrangement of crossing Josephson vortices remains about the same. 

\begin{figure*}[h]
\includegraphics[width=0.9\textwidth]{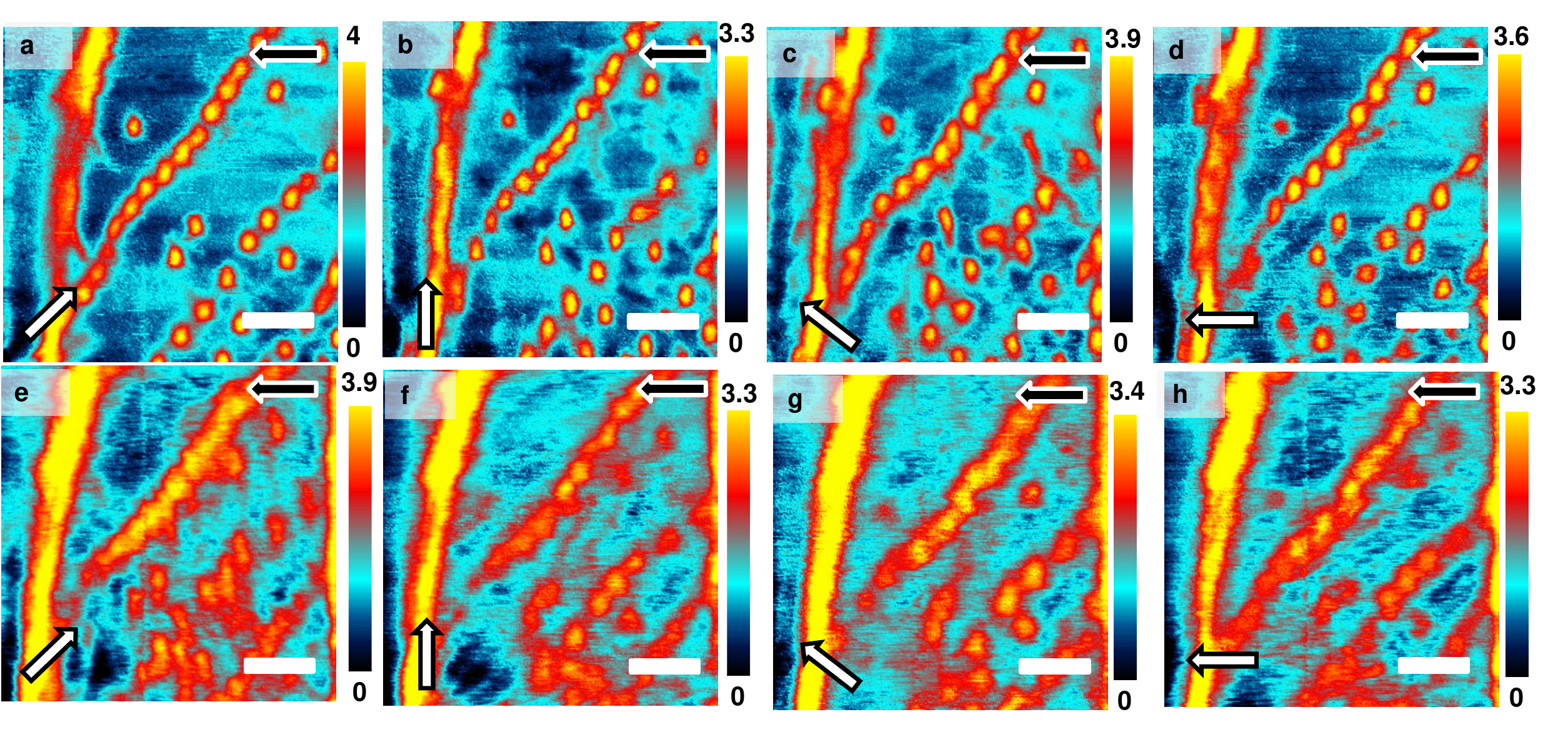} 
\vskip 0cm
\caption{\textbf{Modifying the direction of the in-plane magnetic field in a situation where Josephson vortices cross.} We create a situation where Josephson vortices cross, following the procedure described in the paper, see {\bf a}. The in-plane magnetic field is of 200 G. We then modify the tilt of the magnetic field as shown by the white arrows in each panel. The fast scanning direction is given by the black arrows. Images {\bf a-c} are made at 5.3 K and images {\bf d-f} at 12 K. Color scale provides the changes in the phase of the oscillation of the cantilever and white bars in the images are of
$2.4 \mu m$ size.}
\label{Fig9}
\end{figure*}

\section{Appendix IV. Moving pancake vortices along the tilt of the magnetic field}

In Supplementary Figure\,\ref{Fig10} {\bf a-c} we show vortex drag along the tilt of the magnetic field when there is no crossing. The area shows just a few pancakes, with no clearly discernible pinning at Josephson vortices. At 5\,K, the tip alone is not enough to produce vortex motion. But when we apply a magnetic field perpendicular to the fast scanning direction, the vortex lattice becomes gradually more blurred along the tilt of the magnetic field. When the tilt of the magnetic field and the slow scanning axis coincide (Supplementary Figure\,\ref{Fig10} {\bf c}), pancake vortices are dragged by the tip along the tilt of the magnetic field. Here, the dithering field of the scanning tip reduces pinning and allows for vortex motion, which occurs along the slow scanning direction (or here the tilt of the magnetic field). We use this experiment to estimate the attractive force between the tip and the pancake vortices as $F=\alpha\frac{m\Phi_{0}}{2\pi(z+h_{0})^{2}}$ where $\alpha$ is a numerical constant, $m$ is the dipolar moment per unit length of the tip, $h_{0}$ is an offset in the tip-sample separation and $z$ is the tip-sample
distance\cite{Straver2008,Auslaender2008,doi:10.1142/S0217979210056384,PhysRevB.80.054513}. We consider tip and vortices as magnetic monopoles. We find $F\approx$ 80 pN. The Lorentz force by the Josephson currents acting on pancake vortices can be estimated to be of 50 pN, using the expected value of the currents in a Josephson vortex\cite{PhysRevB.68.094520}. Thus the forces needed to drag pancake vortices are about several tens to hundred pN at 5 K.

\begin{figure*}[h]
\includegraphics[width=0.9\textwidth]{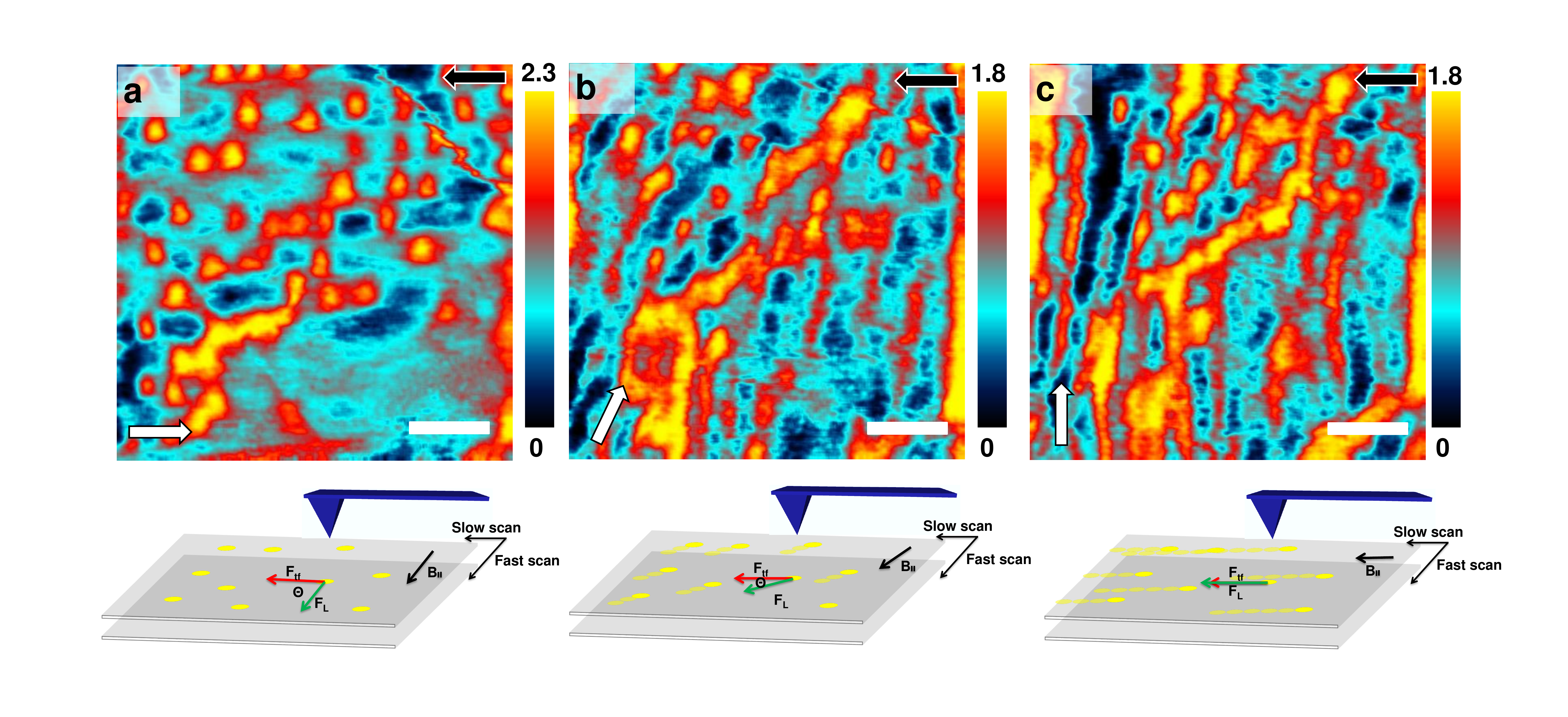}  
\vskip 0cm
\caption{\textbf{Pancake vortices moving along the tilt of the magnetic field.} We show the
evolution of a pancake vortex lattice when modifying the in-plane component of the magnetic field at 5 K. The absolute value of the in-plane magnetic field remains constant at 200 G and is oriented along the white arrows. Black arrows provide the fast scanning direction. Color scale provides the changes in the phase of the oscillation of the cantilever and white bars in the images are of
$2.4 \mu m$ size.  When the tilt of the magnetic field is along the fast scan axis, the Lorentz force ($F_{L}$ in bottom panels) is perpendicular to $F_{tf}$. If the temperature is low enough, $F_{tf}$ is too small to produce vortex motion but vortices can move when the tilt of the magnetic field ($F_{L}$ in bottom panels) coincides with the slow scanning direction and both Lorentz force and the magnetic force from the tip act on the vortices.}
\label{Fig10}
\end{figure*}

\section{Appendix V. Pancake vortex density when modifying the polar tilt}

In the Supplementary Figure\,\ref{Fig11} {\bf a-g} we show the evolution of the vortex lattice when
changing the tilt of the magnetic field from nearly parallel to perpendicular to the layers, but having rapidly ramped up to close to the
critical temperature after each change of the magnetic field. When the field is nearly perpendicular to the sample, the Abrikosov vortex lattice is
hexagonal (Supplementary Figure\,\ref{Fig11} {\bf g}). The density of pancake vortices follows expectations from the surface projection of the magnetic field when the field is tilted (Supplementary Figure\,\ref{Fig11} {\bf h}). We observe linear structures with higher vortex density at all tilts, due to a long pinning center (Supplementary Figure\,{\bf a-g}). It is useful to discuss the apparent vortex size. For instance, we observe in Supplementary Figure\,\ref{Fig11} {\bf d} that pancake vortices lying close to the Josephson vortex on the left side of the image appear  larger than those elsewhere. These vortices are attracted towards the pinning center, so that the corresponding dynamic depinning threshold is reduced and they feel more the oscillatory motion of the MFM tip than their neighbors.

\begin{figure*}[h]
\includegraphics[width=0.95\textwidth]{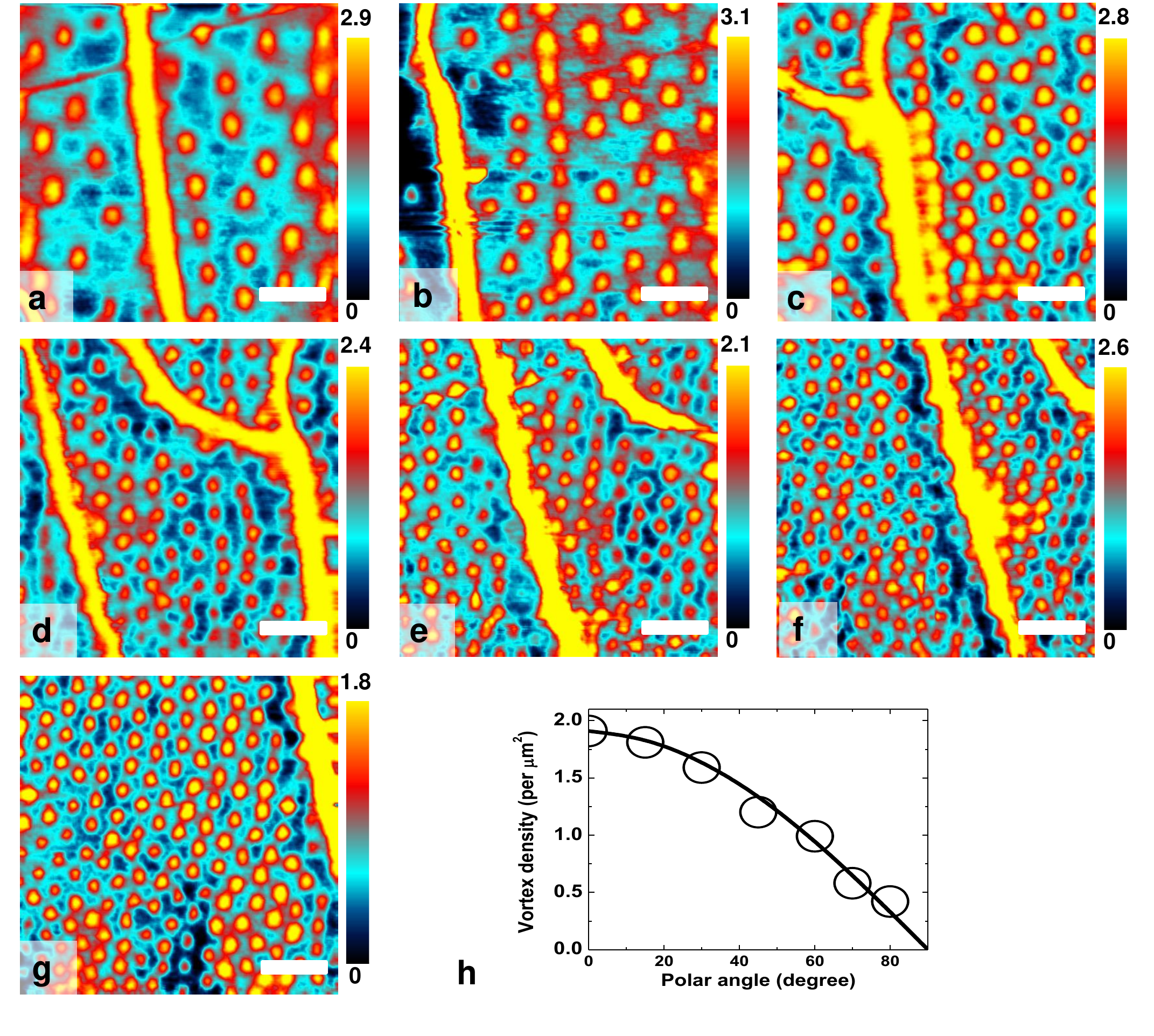}  
\vskip 0cm  
\caption{{\bf Pancake vortex density changes as expected with the magnetic field in field cooled conditions.} MFM images taken at a constant magnetic field of 27.5 G and varying the magnetic field from nearly parallel to nearly perpendicular to the layers {\bf a-g}. All images are taken at 5.3 K in field cooled conditions, after modifying the magnetic field close to the superconducting critical
temperature. Polar angle is {\bf a} 80$^{\circ}$, {\bf b} 70$^{\circ}$, {\bf c}
60$^{\circ}$, {\bf d} 45$^{\circ}$, {\bf e} 30$^{\circ}$, {\bf f} 15$^{\circ}$ and {\bf g} 0$^{\circ}$. White bars in the images are of $2.4 \mu m$ size and color scale bar represents the shift in the phase of the oscillation of the cantilever. The field of view changes slightly at each frame due to a drift produced by heating. In {\bf h} we show the vortex density (taking just pancake vortices in between the stripes of enhanced flux) as a function of the tilt of the magnetic field (points). Line is the density expected from the surface projection of the bulk vortex lattice.}
\label{Fig11}
\end{figure*}

\section{Appendix VI. Vortex size from MFM images and its temperature dependence}

In the Supplementary Figure\,\ref{Fig12} we discuss the shape of vortices versus temperature. As we show in Supplementary Figure\,\ref{Fig12}{\bf a}, the usual shape of pancake vortices has a bell-like form which can be approximated by a Gaussian. When increasing temperature, the shape of the curve is modified. The contrast, as given by the change in phase in degrees, decreases and the curve flattens. We can define an apparent size of the pancake vortex and follow it as a function of temperature (as shown in Fig.\,2 of the publication) and, for a fixed temperature, as a function of the in-plane direction of the magnetic field (Supplementary Figure\,\ref{Fig10}). This value increases with temperature, as shown in Supplementary Figure\,\ref{Fig12}{\bf b}. Vortices become more blurred due to the combined action of the dithering magnetic field of due to the scanning MFM tip and the thermal excitation. When turning the magnetic field, the combined action of the Lorentz force and the dithering magnetic field of the MFM tip (Supplementary Figure\,\ref{Fig10}) add together to produce blurred vortices (Supplementary Figure \ref{Fig12}{\bf b}).

\begin{figure*}[h]
\includegraphics[width=0.95\textwidth]{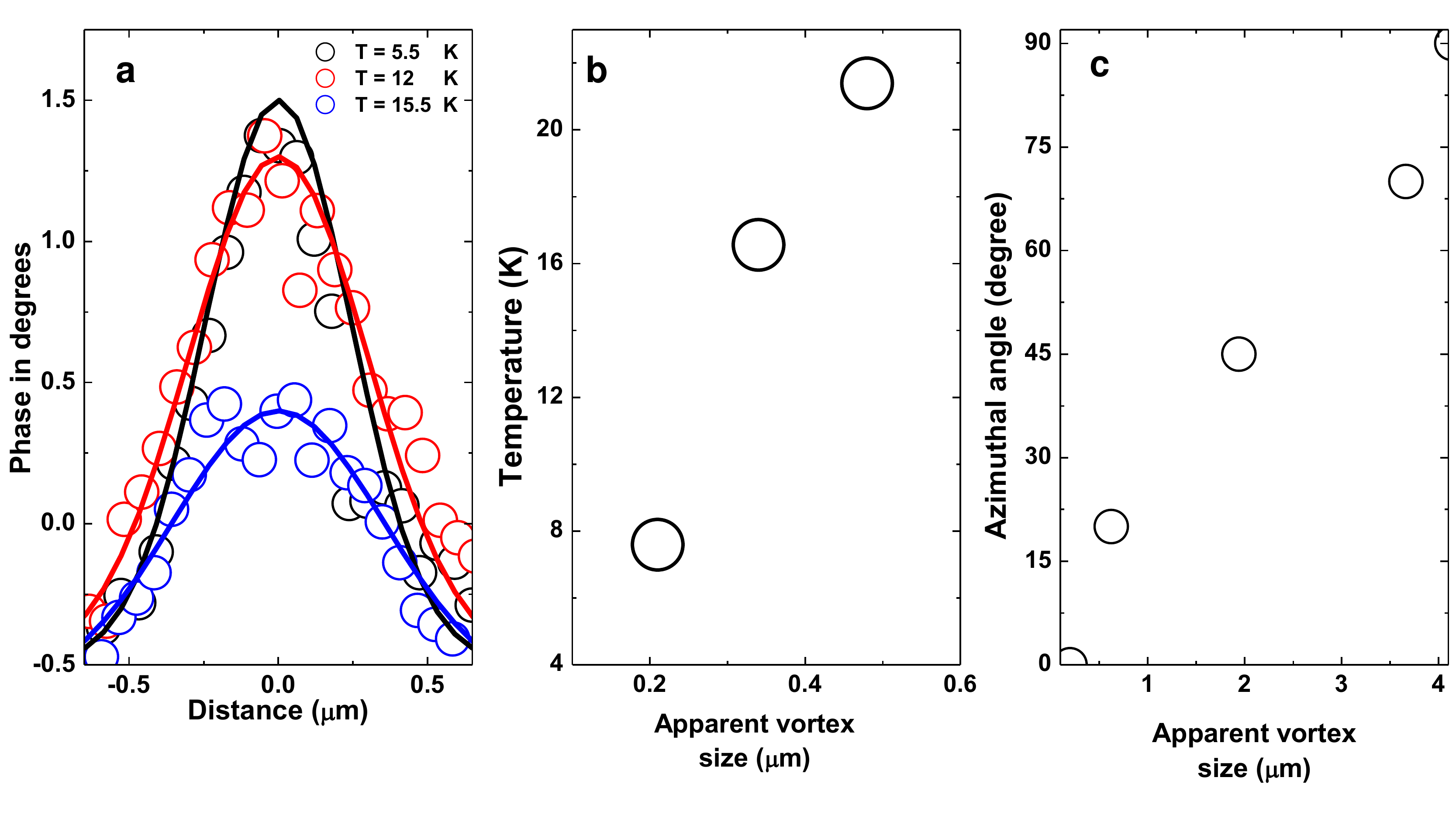} 
\vskip 0cm  
\caption{{\bf Apparent vortex core size as a function of temperature and the in-plane direction of the magnetic field.} {\bf a} We show a line profile across a vortex as a function of temperature, taken from the Fig.\,2 of the main text. Open dots are the line profile and lines are fits to a Gaussian function at each temperature. {\bf b} Apparent vortex size obtained from Gaussian fits to a profile over a single vortex at different temperatures, from the images in the Fig.\,2 of the main text. {\bf c} Apparent vortex size obtained from Gaussian fits to profiles of single vortices from the images shown in Supplementary Figure\,\ref{Fig10}. Notice that, in this case, there is a clear directional motion, which produces elongated shapes that lead, in this plot, to larger sizes.}
\label{Fig12}
\end{figure*}

\section{Appendix VII. Features in the surface that induce strong pinning}

In the Supplementary Figure\,\ref{Fig13} we show a topography image made using the AFM mode together with one of the MFM images of the Figure 4 of the publication. The surface has clearly a very large wrinkle, of an approximate height of 100 nm. This features serves as a strong pinning center of vortices and allows that vortices remain there when modifying the in-plane direction of the magnetic field. This makes it possible to make the vortex manipulation experiment discussed in the Figure 4 of the publication.

\begin{figure*}[h]
\includegraphics[width=0.95\textwidth]{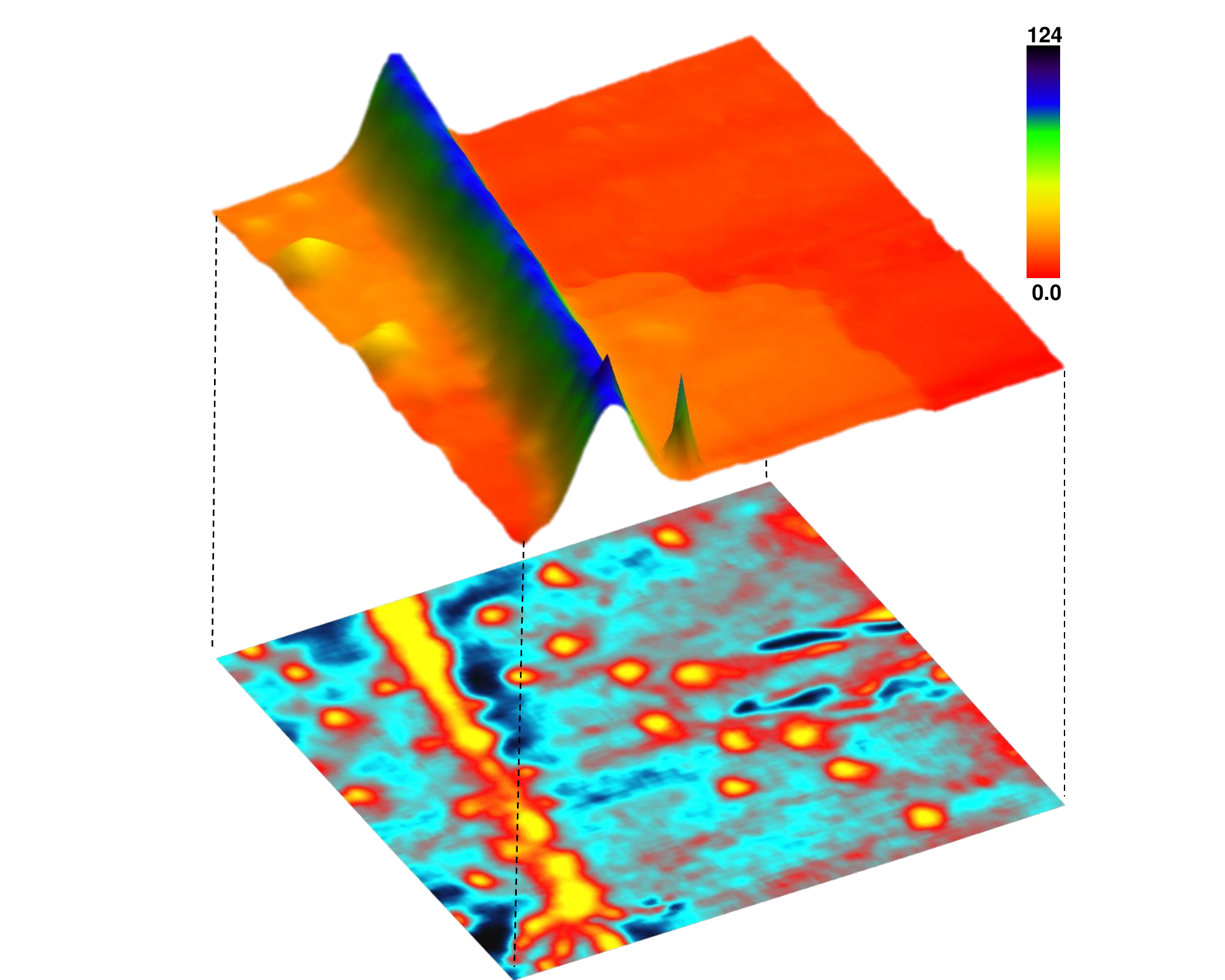} 
\vskip 0cm  
\caption{{\bf Topography of a field of view with one large linear pinning center.} {In the bottom panel we show one of the MFM images of the Figure 4 of the publication, with the long Josephson vortex pinned at the bottom left part. In the upper panel we show the corresponding topography image. The color scale represents the z-axis corrugation of the surface and is given by the bar at the right (in nm).}}
\label{Fig13}
\end{figure*}

\bibliographystyle{naturemag}
%\bibliography{biblio_MFM_BSCCO}

\end{document}